\documentclass[11pt,english]{article}
\usepackage[T1]{fontenc}
\usepackage[latin9]{inputenc}
\usepackage{datetime}
\usepackage{color}
\usepackage{amsthm}
\usepackage{amsmath}
\usepackage{amssymb}
\usepackage{cite}

  \theoremstyle{plain}
  
  \theoremstyle{plain}
  \newtheorem*{lem*}{Lemma}
\theoremstyle{plain}
\newtheorem{thm}{Theorem}
\newtheorem{mydef}{Definition}

\newcommand{\be}{\begin{equation}}
\newcommand{\ee}{\end{equation}}

\newcommand{\bk}[2]{\langle#1|#2\rangle}
\newcommand{\ket}[1]{|#1\rangle}

\newtheorem{corollary}{Corollary}
\newtheorem{prop}{Proposition}

\usepackage{amsfonts}
\usepackage{amsthm}
\usepackage[labelfont=bf, font=small]{caption}
\usepackage{babel}
\begin{document}

%+Title
\title{\large\bf{Isolated Hadamard Matrices from Mutually Unbiased Product Bases}}
\author{Daniel McNulty and Stefan Weigert\\
 Department of Mathematics, University of York\\
York YO10 5DD, United Kingdom\\ \\
\small{\tt{dm575@york.ac.uk, stefan.weigert@york.ac.uk}}}
\date{4 December 2012}
\maketitle

%-Title

%+Abstract
\begin{abstract}
A new construction of complex Hadamard matrices of composite order $d=pq$, with primes $p,q$, is presented which is based on pairs of mutually unbiased bases containing only product states. For product dimensions $d < 100$, we illustrate the method by deriving many previously unknown complex Hadamard matrices. We obtain at least 12 new \emph{isolated} matrices of Butson type, with orders ranging from 9 to 91.
\end{abstract}
%-Abstract

\section{Introduction}

A square matrix $H$ of order $d$ is a complex Hadamard matrix if it is unitary, $HH^\dagger=I$, and if its elements have equal modulus. This definition generalises the concept of a \emph{real} Hadamard matrix with matrix elements limited to the values $\pm 1/\sqrt{d}$. The first known construction of such matrices is due to J.J. Sylvester \cite{sylvester67} while they take their name from J. Hadamard who found that the absolute value of the determinant of a unitary matrix achieves its maximum if all its matrix elements have the same modulus \cite{hadamard93}.

Since then, complex Hadamard matrices have made their appearance in various branches of both mathematics and physics. For example, they relate to the problems of finding bi-unitary sequences and cyclic $n$-roots \cite{bjork91}, they can be useful in constructing certain *-subalgebras of finite von Neumann algebras \cite{popa83}, and error correcting codes \cite{agaian85}. They also have applications in quantum information, representing an important ingredient in teleportation and dense coding schemes \cite{werner01}, and they are closely linked to mutually unbiased bases \cite{schwinger60}. For a detailed overview of their applications, see \cite{agaian85,Horadam07}.

In view of their many uses, a complete classification of complex Hadamard matrices would be highly desirable but has not yet been achieved. All complex Hadamard matrices, up to equivalence, are known for dimensions $d\leq5$ \cite{haagerup97,weigert10}, but their classification remains incomplete for higher dimensions. Many continuous families of complex Hadamard matrices already exist for $d=6$, including the three- and four-parameter families found in \cite{karlsson11} and \cite{szollosi11}, respectively; in addition, a single isolated complex Hadamard matrix, disconnected from any family has been found \cite{tao04,moorhouse01}. A matrix is isolated if its \emph{defect} -- an upper bound on the dimensionality of the set of Hadamard matrices stemming from the matrix -- is zero \cite{tadej06}.

General construction methods exist in composite dimensions \cite{hosoya03, dita04}, and continuous families of complex Hadamard matrices have been obtained from so-called \emph{parameterisations} of known Hadamard matrices \cite{dita04,matolcsi07,dita09,dita10}. There has also been some success in finding continuous families of complex Hadamard matrices for certain prime dimensions \cite{petrescu97}.
A survey of known complex Hadamard matrices is given in \cite{tadej06} for $d\leq16$, with an updated online catalogue provided by \cite{karol12}.

In this paper we introduce a new technique to construct complex Hadamard matrices of order $d=pq$, $p\leq q$, with $p$ and $q$ prime, based on a link between Hadamard matrices and \emph{pairs} of MU bases consisting entirely of \emph{product} states. A $(d \times d)$ complex Hadamard matrix $H$ is unitary, which is another way of saying that its columns represent orthonormal vectors of a basis in a Hilbert space of dimension $d$. By choosing its matrix elements to have modulus $1/\sqrt{d}$, the columns of $H$ will form a basis which is \emph{mutually unbiased} (MU, for short) to the standard basis, since the squared modulus of the inner product of any vector from $H$ with any vector from the standard basis equals $1/d$. In other words, the matrices $\{I,H\}$, where $I$ is the $(d\times d)$ identity matrix, represent a pair of MU bases. Since any pair of MU bases can be brought into \emph{standard} form, which contains the standard basis, the classification problem of complex Hadamard matrices of order $d$ is equivalent to the problem of finding all \emph{pairs} of MU bases in the Hilbert space $\mathbb{C}^d$.

The method we use to construct Hadamard matrices originates from earlier studies of \emph{pairs} of MU \emph{product} bases in dimension six \cite{wiesniak11,mcnulty11,weigert11}. Generalising this method from $d=2\times3$ to composite dimensions $d=pq$, we establish a general construction method resulting in previously undiscovered complex Hadamard matrices. 

The paper is organised as follows: In Section \ref{Sec:CHM} we summarise those properties of complex Hadamard matrices which we will use in later sections. Section \ref{Sec:Construction CHM} includes our first main result, Theorem \ref{thm:newhadamard}, which describes a general construction for complex Hadamard matrices of size $d=pq$. In Section \ref{Sec:Examples of CHM} we apply Theorem \ref{thm:newhadamard} to find new complex Hadamard matrices for $d\leq 15$. We briefly touch upon higher dimensions as well as potential generalisations of the construction in Section \ref{sec:generalisations}. Section \ref{Sec:Summary} contains a summary of our results.

\section{Complex Hadamard matrices \label{Sec:CHM}}

\subsection{MU bases and Hadamard matrices}
 
Two orthonormal bases $\mathcal{B}=\{\ket{\phi_i}\}$ and $\mathcal{B}^\prime=\{\ket{\psi_j}\}$ in a Hilbert space $\mathbb{C}^d$ are MU if and only if $|\bk{\phi_i}{\psi_j}|^2=1/d$ for all $i,j=1\ldots d$. In prime and prime power dimensions $p^n$, $n\in \mathbb{N}$, complete sets of $(p^n+1)$ MU bases exist \cite{wootters89,ivanovic81}. In \emph{composite} dimensions such as $d=6,10,12,...$, the question of whether complete sets of $(d+1)$ MU bases exist remains open \cite{durt10}.

MU bases in a finite-dimensional Hilbert space $\mathbb{C}^d$ are closely related to complex Hadamard matrices. Given a set of $(r+1)$ MU bases in standard form $\{\mathcal{I},\mathcal{B}_1,\ldots\mathcal{B}_r\}$, with $\mathcal{I}$ the standard basis, the bases $\mathcal{B}_1,\ldots,\mathcal{B}_r$ are represented by $(d\times d)$ complex Hadamard matrices, $H_1,\ldots,H_r$. Here the vectors of each basis are given by the columns of the respective matrix. Since these matrices are MU to the identity matrix, their matrix elements have modulus $1/\sqrt{d}$.

Two complex Hadamard matrices $H$ and $K$ are \emph{equivalent} to each other if one can write $H=D_1P_1KP_2D_2$, where $D_1,D_2$ are diagonal matrices with diagonal elements consisting of phase factors and $P_1,P_2$ are permutation matrices. As a consequence, every complex Hadamard matrix can be \emph{dephased} such that the matrix elements of its first row and column equal $1/\sqrt{d}$.

A useful criterion to determine whether two complex Hadamard matrices are \emph{inequivalent} is to calculate their \emph{Haagerup set}, defined in \cite{haagerup97}.
\begin{mydef}
The Haagerup set of a complex Hadamard matrix $H$ of order $n$ with matrix elements $h_{ij}$ is given by
\begin{equation}
\Lambda(H)=\{h_{ij}h_{kl}\bar{h}_{il}\bar{h}_{kj}:i,j,k,l=1,\ldots,n\},
\end{equation}
where $\bar{h}_{ij}$ denotes the complex conjugate of $h_{ij}$.
\end{mydef}
Two matrices with \emph{different} Haagerup sets are \emph{inequivalent} since the set $\Lambda(H)$ is invariant under equivalences. Note, however, that two inequivalent matrices may have the same Haagerup set.

Sets of MU bases are \emph{equivalent} if they can be mapped to each other by any of the following unitary or anti-unitary equivalence transformations: a unitary map acting on all bases within a set; the multiplication of states within bases by arbitrary phase factors; the permutation of states within a basis, and the complex conjugation of all bases. In addition, one may exchange any two bases within a set.

It is often the case that a single Hadamard matrix is contained in a subset of a larger family of complex Hadamard matrices of which there are two types: \emph{affine} and \emph{non-affine} families. The following definition, given in \cite{tadej06}, uses the notation $\circ$ as the entrywise product of matrices (also known as their Hadamard product), and EXP$(.)$ denotes the entrywise exponential function acting on a matrix.
\begin{mydef}
Given a $(d\times d)$ complex Hadamard matrix $H$, an affine family of Hadamard  matrices stemming from it is given by 
\begin{equation}
H(\mathcal{R})=\{H\circ \text{\emph{EXP}}(i\cdot R) : R\in \mathcal{R}\},
\end{equation}
where $\mathcal{R}$ is a subspace of all real $(d \times d)$ matrices with zeros in the first row and column.
\end{mydef}
A family of Hadamard matrices is called \emph{non-affine} if it cannot be written in this form.

The \emph{defect} $d(H)$ of a Hadamard matrix $H$, defined in \cite{tadej06}, provides an upper bound on the dimensionality of any set of Hadamard matrices stemming from $H$. If a dephased Hadamard matrix has a defect of zero then the matrix is called \emph{isolated}, expressing the fact that all complex Hadamard matrices in a neighbourhood of $H$ are equivalent. To calculate the defect of $H$, we multiply all elements of the Hadamard matrix with independent phase factors, i.e. $H_{ij}\rightarrow e^{a_{ij}}H_{ij}$ for $i,j=2\ldots d$, and solve the set of equations, to first order, which are imposed by the unitarity condition (see \cite{bengtsson07} for an explicit example). For all but the smallest dimensions $d$ or special cases, it seems imperative to use a computer program in order to determine the defect; the software we have used is MATLAB \cite{Matlab}. The defect provides only a weak upper bound on the dimensionality of a Hadamard family; higher-order solutions of the unitarity conditions often lead to stronger bounds \cite{sa12}.

\subsection{Known constructions in composite dimensions}

There are many known constructions of complex Hadamard matrices (cf. \cite{karol12}) some of which apply only to specific dimensions. We briefly review two constructions of (affine) complex Hadamard matrices based on the tensor product of smaller matrices following \cite{szollosi12}.

\begin{thm}\emph{(Hosoya-Suzuki \cite{hosoya03})}
Let $M_1, M_2,\ldots,M_v$ be $k \times k$, $N_1, N_2,\ldots,N_k$ be $v \times v$ complex Hadamard matrices. Then the generalised tensor product matrix, denoted by $Q=(M_1,M_2,\ldots,M_v)\otimes(N_1,N_2,\ldots,N_k)$, whose $(i,j)$th block is given by the matrix $Q_{ij}=diag([M_1]_{ij},[M_2]_{ij},\ldots,[M_v]_{ij})N_j$, is a complex Hadamard matrix of order $vk$.
\end{thm}
By using a simpler version of this tensor product structure, these matrices can be \emph{parameterised}, i.e. embedded in larger families of Hadamard matrices.
\begin{corollary}\emph{(Di\c{t}\v{a} \cite{dita04})}\label{thm:dita}
Let $M=(m_{ij})$ be a $k\times k$ and $N_1,N_2,\ldots,N_k$ be $v \times v$ dephased complex Hadamard matrices with $m$ and $n_1,n_2,\ldots,n_k$ free parameters, respectively. Then the block matrix
\begin{equation}
Q=\left(\begin{array}{cccc}
m_{11}N_1 & m_{12}N_2 & \ldots & m_{1k}N_k\\
\vdots &  &  & \\
m_{k1}N_1 & m_{k2}N_2 & \ldots & m_{kk}N_k\end{array}\right),
\end{equation}
with blocks $Q_{ij}=m_{ij}N_j$, is a complex Hadamard matrix of order $vk$ with $m+\sum_{i=1}^{k}n_i+(k-1)(v-1)$ free parameters.
\end{corollary}
Any matrix which can be derived from Corollary \ref{thm:dita} is called a \emph{Di\c{t}\v{a}-type} complex Hadamard matrix.

\subsection{Butson-type complex Hadamard matrices}

We finally recall a special class of Hadamard matrices called \emph{Butson-Hadamard} matrices. 
\begin{mydef}
A complex Hadamard matrix of order $d$ is a Butson-Hadamard matrix $BH(d,r)$ if its elements are $r^{th}$ roots of unity, apart from a factor $1/\sqrt{d}$.
\end{mydef}

Note that some authors write $BH(r,d)$ instead of $BH(d,r)$. It is straightforward to show that a Hadamard matrix is (equivalent to one) of Butson-type $BH(d,r)$: once dephased, all its matrix elements must be $r^{\mbox{\tiny{th}}}$ roots of unity.

The simplest examples of Butson-type matrices occur when $r=2$; in this case the matrices $BH(d,2)$ are the set of $(d\times d)$ \emph{real} Hadamard matrices. The existence of $BH(d,r)$ matrices for arbitrary values of $d$ and $r$ is still an open problem; it remains unknown, for example, if real Hadamard matrices of the form $B(4n,2)$ exist for all integers $n$. A summary of existing Butson-Hadamard matrices with fourth and sixth roots of unity can be found in \cite{szollosi12} and the known $BH(d,r)$ matrices for $d \leq 16$ are given in \cite{karol12}. There are also several existence theorems for $BH(d,r)$ matrices, e.g.

\begin{thm}\label{thm:butson}\emph{(Butson \cite{butson62})}
When $p$ is prime, a $BH(2p,p)$ matrix can be constructed.
\end{thm}
 
For $p=3$, the matrix $BH(6,3)$ turns out to be the isolated matrix $S_6$, which was also found independently in \cite{tao04,moorhouse01}. We have derived the matrices $B_{10}\in BH(10,5)$ and $B_{14}\in BH(14,7)$ following Butson's method (see Appendix \ref{Butson's construction}) since they will be important in the present context and seem to be unavailable in the literature. 

Butson-Hadamard matrices will appear in Sections \ref{Sec:Examples of CHM} and \ref{sec:generalisations}, where we will derive previously unknown examples of complex Hadamard matrices of orders up to 91. Most of these examples cannot be constructed from Theorem \ref{thm:butson}.

\section{Complex Hadamard matrices from pairs of MU product bases \label{Sec:Construction CHM}}

The following theorem shows how to construct a complex Hadamard matrix of order $d=pq$ where $p$ and $q$ are both prime, using sets of MU bases in dimensions $p$ and $q$.

\begin{thm} \label{thm:newhadamard}Suppose that $K_0,\ldots,K_{p-1}$ and $L_0,\ldots,L_{p-1}$ are unitary matrices of order $q$ such that $K^\dagger_mL_n$ are complex Hadamard matrices for all $m,n=0,\ldots,p-1$, i.e. $K_m$ is MU to $L_n$, and let $\alpha_{ij}/\sqrt{p}$ be the $(i,j)$th element of a complex Hadamard matrix $M$ of order $p$, with $|\alpha_{ij}|=1$. Then the block matrix $H_{pq}$ given by
\begin{equation}\label{generalhadamard}
H_{pq}=\frac{1}{\sqrt{p}}\left(\begin{array}{ccccc}
\alpha_{11}K_0^\dagger L_0 & \alpha_{12}K_0^\dagger L_1 & \ldots & \alpha_{1p}K_0^\dagger L_{p-1}\\
\alpha_{21}K_1^{\dagger}L_0 & \alpha_{22}K_1^{\dagger}L_1 & \ldots & \alpha_{2p}K_1^{\dagger}L_{p-1}\\
\alpha_{31}K_2^{\dagger}L_0 & \alpha_{32}K_2^{\dagger}L_1 & \ldots & \alpha_{3p}K_2^{\dagger}L_{p-1}\\
\vdots &  &  & \\
\alpha_{p1}K_{p-1}^{\dagger}L_0 & \alpha_{p2}K_{p-1}^{\dagger}L_1 & \ldots & \alpha_{pp}K_{p-1}^{\dagger}L_{p-1}\end{array}\right)
\end{equation}
is a complex Hadamard matrix of order $pq$.
\end{thm}

The theorem follows easily from factorising the matrix $H_{pq}$ such that $H_{pq}=B^\dagger_1B_2$, where
\begin{equation}
B_1=\left(\begin{array}{cccc}
K_0 & 0 & \ldots & 0\\
0 & K_1 & \ldots & 0\\
\vdots &\vdots &\ddots &\vdots\\
0 & 0 & \ldots & K_{p-1}\end{array}\right)
\end{equation}
and
\begin{equation}
B_2=\left(\begin{array}{cccc}
\alpha_{11}L_0 & \alpha_{12}L_1 & \ldots & \alpha_{1p}L_{p-1}\\
\alpha_{21}L_0 & \alpha_{22}L_1 & \ldots & \alpha_{2p}L_{p-1}\\
\vdots & \vdots &  & \vdots\\
\alpha_{p1}L_0 & \alpha_{p2}L_1 & \ldots & \alpha_{pp}L_{p-1}\end{array}\right).
\end{equation}
The column vectors of the unitary matrices $B_1$ and $B_2$ form a pair of MU bases since the block matrices $K_m$ are MU to $L_n$, i.e. $K^\dagger_m L_n$ are complex Hadamard matrices for all $m,n=0,\ldots,p-1$. Thus, by mapping $B_1$ to the identity matrix using the unitary transformation $B_1^{\dagger}$,  the matrix $B_2$ is simultaneously mapped to $B_1^\dagger B_2=H_{pq}$, which is a complex Hadamard matrix. This completes the proof of Theorem \ref{thm:newhadamard}.

In fact, the matrices $B_1$ and $B_2$ correspond to a pair of MU \emph{product} bases where the columns of $B_1$ and $B_2$ form the vectors of each basis.
We can write the pair of matrices $B_1$ and $B_2$ as the orthonormal bases
\begin{equation}\label{basis1}
\mathcal{B}_1=\Bigl\{\ket{0_z}\otimes\mathcal{K}_0,\,\ket{1_z}\otimes\mathcal{K}_1,\,\ldots,\,\ket{(p-1)_z}\otimes\mathcal{K}_{p-1}\Bigr\}
\end{equation}
and
\begin{equation}\label{basis2}
\mathcal{B}_2=\Bigl\{\ket{0_a}\otimes\mathcal{L}_0,\,\ket{1_a}\otimes\mathcal{L}_1,\,\ldots,\,\ket{(p-1)_a}\otimes\mathcal{L}_{p-1}\Bigr\}\,,
\end{equation}
respectively, where $\ket{m_z}\otimes\mathcal{K}_m$ denotes the tensor product of a state $\ket{m_z}$ from the standard basis of $\mathbb{C}^p$ with all states from a basis $\mathcal{K}_m$ of the space $\mathbb{C}^q$ corresponding to the matrix $K_m$. Similarly, $\ket{n_a}\otimes \mathcal{L}_n$ is defined such that $\ket{n_a}$ is a state in $\mathbb{C}^p$ corresponding to the $n$th column vector of the $(p\times p)$ Hadamard matrix $M$, and $\mathcal{L}_n$ is a basis of $\mathbb{C}^q$ corresponding to the matrix $L_n$. Thus, by mapping $\mathcal{B}_1$ to the standard basis, the columns of the second basis $\mathcal{B}_2$ form the complex Hadamard matrix $H_{pq}$.

To simplify the matrix $H_{pq}$ we perform equivalence transformations on the pair of MU bases \{$\mathcal{B}_1, \mathcal{B}_2\}$ such that $\mathcal{K}_0$ and $\mathcal{L}_0$ are mapped to the standard and Fourier basis of $\mathbb{C}^q$ respectively, and the orthonormal basis $\{\ket{0_a},\ldots,\ket{(p-1)_a}\}$ is mapped to the Fourier basis of $\mathbb{C}^p$, i.e. $K_0\equiv I_q$,  $L_0\equiv F_q$ and $M\equiv F_p$, with $I_q$ the $(q\times q)$ identity matrix, and $F_p$, $F_q$, being the Fourier matrices of order $p$ and $q$ respectively. Since $B_1$ is MU to $B_2$, the set $\{I_q,K_1,K_2,\ldots,K_{p-1}\}$ is MU to $\{F_q,L_1,\ldots,L_{p-1}\}$, and as a consequence, $L_1,\ldots,L_{p-1}$ are complex Hadamard matrices. We will continue to use the simplification $M\equiv F_p$, $K_0\equiv I_q$ and $L_0\equiv F_q$ throughout.
 
In the trivial case of $K_1=\ldots=K_{p-1}=I_q$, one can choose each matrix $L_n$ to be a $(q-1)$-parameter family $DF_q$ where $D=\mbox{diag}(1,e^{a^n_1},\ldots,e^{a^n_{q-1}})$ for each $n>0$. In this case, $H_{pq}$ is a $(p-1)(q-1)$-parameter family of complex Hadamard matrices.

In the following section we will show that for certain choices of the $(q\times q)$ matrices $K_1,\ldots,K_{p-1},L_1,\ldots,L_{p-1}$, the matrix $H_{pq}$ given in Theorem \ref{thm:newhadamard} supplies new examples of Hadamard matrices.  Most of the matrices we find will be isolated, which is sufficient to confirm that Theorem \ref{thm:newhadamard} produces matrices not of Di\c{t}\v{a}-type: every Di\c{t}\v{a}-type matrix is embedded within a family depending on at least $(k-1)(v-1)$ free parameters, with $k,v>1$. 

\section{Examples: $d\leq 15$ \label{Sec:Examples of CHM}}

We will now use the construction given in Theorem \ref{thm:newhadamard} to find complex Hadamard matrices of composite dimensions $d=pq$, with prime numbers $p\leq q$. In this section, we limit ourselves to matrices of order $d \leq 15$ with $p \leq q$. Larger dimensions and possible generalisations of the construction will be considered briefly in Sec. \ref{sec:generalisations}.

\subsection{Dimension four}

In dimension four, all inequivalent complex Hadamard matrices are given by the one-parameter family $F_4(a)$, $a\in[0,\pi]$ \cite{haagerup97,weigert10}. We re-derive this family from Theorem \ref{thm:newhadamard} using the block matrix,
\begin{equation}\label{hadamardd=6}
H_4=\frac{1}{\sqrt{2}}\left(\begin{array}{cc}
F_2 & L_1\\
K_1^{\dagger}F_2 & -K_1^{\dagger}L_1\end{array}\right),
\end{equation}
where
\begin{equation}\label{fouriermatrixd=3}
F_2\equiv\frac{1}{\sqrt{2}}\left(\begin{array}{cc}
1 & 1\\
1 & -1\end{array}\right)
\end{equation}
is the $(2\times 2)$ Fourier matrix and $L_1, K_1$ are specific unitary matrices of order two: to apply Theorem \ref{thm:newhadamard} it is necessary that the set $\{I_2,K_1\}$ is MU to $\{F_2,L_1\}$.

If $K_1$ is chosen as the identity, then $L_1$ can take the form
\begin{equation}\label{fouriermatrixd=3}
L_1(a)=\frac{1}{\sqrt{2}}\left(\begin{array}{cc}
1 & 1\\
e^{ia} & -e^{ia}\end{array}\right),
\end{equation}
where the column vectors of $L_1$ are indeed MU to the standard basis; an overall phase factor has been removed using equivalence transformations.
 Thus, $H_4$ turns into a one-parameter family of complex Hadamard matrices,
 \begin{equation}\label{hadamardd=6}
H_4(a)=\frac{1}{\sqrt{2}}\left(\begin{array}{cc}
F_2 & L_1(a)\\
F_2 & -L_1(a)\end{array}\right).
\end{equation}
By permuting rows it is easily shown that $H_4(a)$ is equivalent to the one-parameter Fourier family $F_4(a)$. One can exchange $K_1$ with $L_1$ but the resulting family is still equivalent to $F_4(a)$; no other choices are possible. Note that in the four-dimensional case, $F_4(a)$ is equivalent to the transposed Fourier family $(F_4(a))^T$, a relation that does not always hold for larger composite dimensions.

\subsection{Dimension six \label{d=6}}

In a recent paper \cite{weigert11}, pairs of MU product bases of the form $\mathcal{B}_1=\{\ket{0}\otimes\mathcal{K}_0,\ket{1}\otimes\mathcal{K}_1\}$ and $\mathcal{B}_2=\{\ket{0_x}\otimes\mathcal{L}_0,\ket{1_x}\otimes\mathcal{L}_1\}$ were shown to give rise to the transposed Fourier family of complex Hadamard matrices \cite{tadej06} and the isolated matrix $S_6\in BH(6,3)$ \cite{butson62,moorhouse01,tao04}. Here we rederive this result on the basis of Theorem \ref{thm:newhadamard} where we start with the following $(6\times6)$ matrix
\begin{equation}\label{hadamardd=6}
H_6=\frac{1}{\sqrt{2}}\left(\begin{array}{cc}
F_3 & L_1\\
K_1^{\dagger}F_3 & -K_1^{\dagger}L_1\end{array}\right).
\end{equation}
The ($3\times 3$) Fourier matrix is given by
\begin{equation}\label{fouriermatrixd=3}
F_3\equiv\frac{1}{\sqrt{3}}\left(\begin{array}{ccc}
1 & 1 & 1\\
1 & \omega & \omega^{2}\\
1 & \omega^{2} & \omega\end{array}\right),
\end{equation}
with $\omega=e^{2\pi i/3}$ being a third root of unity, and $K_1,L_1$ are unitary matrices of order three. The bases $\mathcal{B}_1$ and $\mathcal{B}_2$ will be MU if the pair $\{I_3,K_1\}$ is MU to the pair $\{F_3,L_1\}$. A proof given in \cite{mcnulty11} limits the possible choices for the matrices $K_1$ and $L_1$ to just three: (i) $K_1=I_3$; (ii) $L_1=F_3$; (iii) all four matrices are pairwise MU.

If $K_1=I_3$, the most general set of matrices satisfying the MU conditions is a two-parameter set, 
\begin{equation}\label{fouriermatrixd=3}
L_1(a,b)=\frac{1}{\sqrt{3}}\left(\begin{array}{ccc}
1 & 1 & 1\\
e^{ia} & \omega e^{ia} & \omega^{2}e^{ia}\\
e^{ib} & \omega^{2}e^{ib} & \omega e^{ib}\end{array}\right).
\end{equation}
Thus, $H_6$ becomes a two-parameter family of complex Hadamard matrices which is equivalent to the transposed Fourier family $(F_6^{(2)})^T$.

If all four matrices $I_3, F_3, K_1$ and $L_1$ are MU then $K_1$ and $L_1$ are given by
\begin{equation}\label{MUbasesd=3}
K_1=H_y\equiv\frac{1}{\sqrt{3}}\left(\begin{array}{ccc}
1 & 1 & 1\\
\omega & \omega^{2} & 1\\
\omega & 1 & \omega^{2}\end{array}\right)\quad \text{and}\quad
L_1=H_w\equiv\frac{1}{\sqrt{3}}\left(\begin{array}{ccc}
1 & 1 & 1\\
\omega^{2} & 1 & \omega\\
\omega^{2} & \omega & 1\end{array}\right),
\end{equation}
or vice versa. Here, $\{I_3,F_3,H_y,H_w\}$ is the complete set of MU bases in $\mathbb{C}^3$. The complex Hadamard matrix $H_6$ in ($\ref{hadamardd=6}$) associated with $K_1=H_y$ and $L_1=H_w$ is equivalent to $S_6$, the only known isolated complex Hadamard matrix of order six.

Thus, we have indeed constructed the transposed Fourier family of complex Hadamard matrices and the isolated matrix $S_6$ from Theorem \ref{thm:newhadamard}. There is a multitude of additional Hadamard matrices of order six, including three- and four-parameter families \cite{karlsson11,szollosi11}, but none of these can be derived from Theorem \ref{thm:newhadamard}.

\subsection{Dimension nine \label{d=9}}

The online catalogue \cite{karol12} lists three types of complex Hadamard matrices of order nine: the four-parameter Fourier family $F_9^{(4)}$, the isolated matrix $N_9$ \cite{beauchamp06}, and the matrix $B_9$ which has a defect of two \cite{beauchamp06}. Three Butson-Hadamard matrices are among them, namely $F_3\otimes F_3 \in BH(9,3)$, $F_9\in BH(9,9)$ and $B_9\in BH(9,10)$. Theorem \ref{thm:newhadamard} allows us to identify an additional isolated Butson-Hadamard matrix of the form $BH(9,6)$.

The matrix in Eq. ($\ref{generalhadamard}$), for $d=9$, has the structure
\begin{equation}
H_{9}=\frac{1}{\sqrt{3}}\left(\begin{array}{ccc}
F_3 & L_1 & L_2\\
K_1^{\dagger}F_3 & \omega K_1^{\dagger}L_1 & \omega^2K_1^{\dagger}L_2\\
K_2^{\dagger}F_3 & \omega^2K_2^{\dagger}L_1 & \omega K_2^{\dagger}L_2\end{array}\right),
\end{equation}
where $\omega=e^{2\pi i/3}$ is a third root of unity, and the matrices $I_3$ and $F_3$ are the $(3 \times 3)$ identity and Fourier matrix, respectively. The $(3 \times 3)$ matrices $K_1, K_2, L_1$ and $L_2$ must be chosen such that the set $\{I_3,K_1,K_2\}$ is MU to $\{F_3,L_1,L_2\}$.

In the six-dimensional case (cf. Sec \ref{d=6}), the choice of pairs $\{I,K_1\}$ and $\{F_3,L_1\}$ was limited to either two matrices having identical columns (up to column permutations) or all four matrices being MU. Similarly, the choices for the triples $\{I_3,K_1,K_2\}$ and $\{F_3,L_1,L_2\}$ is restricted to the following two possibilities: (i) three matrices within one triple are identical (up to column permutations); (ii) two matrices are identical in each triple and all four MU bases $\{I_3,F_3,H_y,H_w\}$ are used.

If all matrices in one triple are the same, i.e. $K_1=K_2=I_3$, then $H_9$ is equivalent to the transposed Fourier family $(F_9^{(4)})^T$ of order $9$, which depends on four real parameters.

Now suppose that $K_1=I_3$ and $L_1=F_3$. The only remaining choice for matrices $K_2$ and $L_2$, (if $K_2\neq I_3$), that satisfy the MU conditions is $K_2=H_y$ and $L_2=H_w$ (or vice versa), with $H_y$ and $H_w$ defined in Eq. ($\ref{MUbasesd=3}$). Denoting the resulting matrix by $S_9$, we find
\begin{equation}\label{d=9matrix}
S_{9}=\frac{1}{\sqrt{3}}\left(\begin{array}{ccc}
F_3 & F_3 & H_w\\
F_3 & \omega F_3 & \omega^2H_w\\
H_y^{\dagger}F_3 & \omega^2H_y^{\dagger}F_3 & \omega H_y^{\dagger}H_w\end{array}\right).
\end{equation}

After dephasing, the matrix $S_9$ takes the form
 
\begin{equation}\label{dephasedmatrixd=9}
S_9=\frac{1}{3}\left(\begin{array}{ccccccccc}
1 & 1 & 1 & 1 & 1 & 1 & 1 & 1 & 1\\
1 & \omega & \omega^{2} & 1 & \omega & \omega^2 & \omega^2 & 1 & \omega\\
1 & \omega^{2} & \omega & 1 & \omega^2 & \omega & \omega^2 & \omega & 1\\
1 & 1 & 1 & \omega & \omega & \omega & \omega^2 & \omega^2 & \omega^2\\
1 & \omega & \omega^2 & \omega & \omega^2 & 1 & \omega & \omega^2 & 1\\
1 & \omega^2 & \omega & \omega & 1 & \omega^2 & \omega & 1 & \omega^2\\
1 & \omega & \omega & \omega^2 & 1 & 1 & -\omega & -1 & -1\\
1 & \omega^2 & 1 & \omega^2 & \omega & \omega^2 & -\omega^2 & -1 & -\omega^2\\
1 & 1 & \omega^2 & \omega^2 & \omega^2 & \omega & -\omega^2 & -\omega^2 & -1\end{array}\right),
\end{equation}
\normalsize
where $\omega=e^{2\pi i/3}$ is a \emph{third} root of unity. Due to the negative signs in the bottom right block, $S_9$ is a Butson-Hadamard matrix containing \emph{sixth} roots of unity, i.e. $S_9\in BH(9,6)$. We find the defect of this matrix to be $d(S_9)=0$ implying that $S_9$ is \emph{isolated}.

\begin{prop}
The matrix $S_9$ is inequivalent to $F_9^{(4)}$, $B_9$ and $N_9$. 
\end{prop}
Since $F_9^{(6)}$ and $B_9$ have non-zero defects and $N_9$ contains only tenth roots of unity, it is clear that $S_9$ is inequivalent to any known complex Hadamard matrix in dimension $d=9$. The only other isolated complex Hadamard matrix known for $d=9$, i.e. $N_9$, was found by a numerical search in \cite{beauchamp06}. As far as we are aware, the matrix $S_9$ has not been published previously.

\subsection{Dimension ten}

The known Hadamard matrices in dimension ten include the Fourier family $F^{(4)}_{10}$ and its transpose $(F^{(4)}_{10})^T$, the family $D^{(8)}_{10}$ found by Di\c{t}\v{a} \cite{dita09} and a family $D^{(3)}_{10}$ stemming from $D_{10}$ \cite{szollosi08}. There is also an isolated matrix $N_{10A}$, a family $N_{10B}^{(3)}$ found originally in \cite{beauchamp06} and parameterised in \cite{lampio12}, and $G_{10}^{(1)}$ \cite{lampio12}. Furthermore, there are the Butson-Hadamard matrices $X_{10}\in BH(10,5)$ \cite{banica07} and $W'\in BH(10,6)$ \cite{compton12}. Within the continuous families, several Butson-type matrices exist: $D_{10}\in BH(10,4)$, $F_2\otimes F_5 \simeq F_{10} \in BH(10,10)$, and those contained in $D^{(8)}_{10}$, e.g. $H^{(\omega)}_{10}$, $d^{(\omega)}_{10}\in BH(10,6)$ given in \cite{dita09}. 

In the following, we construct a complex Hadamard matrix of Butson-type based on the block matrix
\begin{equation}
H_{10}=\frac{1}{\sqrt{2}}\left(\begin{array}{cc}
F_5 & L_1\\
K_1^{\dagger}F_5 & -K_1^{\dagger}L_1\end{array}\right),
\end{equation}
where it is necessary that the pair of $(5\times 5)$ matrices $\{I_5,K_1\}$ is MU to the pair of $(5\times5)$ matrices $\{F_5,L_1\}$. Here, $I_5$ is the identity matrix and $F_5$ the Fourier matrix,

\begin{equation}\label{fourierd=5}
F_5\equiv\frac{1}{\sqrt{5}}\left(\begin{array}{ccccc}
1 & 1 & 1 & 1 & 1\\
1 & \omega & \omega^{2} & \omega^3 & \omega^4\\
1 & \omega^{2} & \omega^4 & \omega & \omega^3\\
1 & \omega^3 & \omega & \omega^4 & \omega^2\\
1 & \omega^4 & \omega^3 & \omega^2 & \omega\end{array}\right)\,,
\end{equation}

with $\omega=e^{2\pi i/5}$ a fifth root of unity. Assuming that $K_1$ and $L_1$ are not identical to $I_5$ and $F_5$, respectively, one choice is to require that the matrices within the set $\{I_5,F_5,K_1,L_1\}$ are pairwise MU.

In dimension five, the complete set of six MU bases can be written as 
\begin{equation}
\{I_5,F_5,H_1,H_2,H_3,H_4\},
\end{equation}
where $H_i$ are the complex Hadamard matrices of order five given by
\begin{equation}\label{MUcompletesetd=5}
H_1=DF_5,\quad H_2=D^2F_5,\quad H_3=D^3F_5,\quad H_4=D^4F_5,
\end{equation}
and with a diagonal matrix,
\begin{equation}
D=\text{diag}(1,\omega,\omega^4,\omega^4,\omega).
\end{equation}
This characterisation of the complete set of MU bases is based on a construction in \cite{weigert10}. By choosing $K_1=H_3$ and $L_1=H_4$, the matrix $H_{10}$ becomes
\begin{equation}\label{d=10matrix}
S_{10}=\frac{1}{\sqrt{2}}\left(\begin{array}{cc}
F_5 & H_4\\
H_3^{\dagger}F_5 & -H_3^{\dagger}H_4\end{array}\right),
\end{equation}
which has the dephased form
\begin{equation}
S_{10}=\frac{1}{\sqrt{10}}\left(\begin{array}{cccccccccc}
1 & 1 & 1 & 1 & 1 & 1 & 1 & 1 & 1 & 1\\
1 & \omega & \omega^{2} & \omega^3 & \omega^4 & \omega^4 & 1 & \omega & \omega^2 & \omega^3\\
1 & \omega^{2} & \omega^4 & \omega & \omega^3 & \omega & \omega^3 & 1 & \omega^2 & \omega^4\\
1 & \omega^3 & \omega & \omega^4 & \omega^2 & \omega & \omega^4 & \omega^2 & 1 & \omega^3\\
1 & \omega^4 & \omega^3 & \omega^2 & \omega & \omega^4 & \omega^3 & \omega^2 & \omega & 1\\
1 & \omega^3 & \omega^2 & \omega^2 & \omega^3 & 1 & \omega & \omega^4 & \omega^4 & \omega\\
1 & \omega^2 & 1 & \omega^4 & \omega^4 & \omega^3 & \omega^2 & \omega^3 & \omega & \omega\\
1 & \omega & \omega^3 & \omega & 1 & \omega^2 & \omega^4 & \omega^3 & \omega^4 & \omega^2\\
1 & 1 & \omega & \omega^3 & \omega & \omega^2 & \omega^2 & \omega^4 & \omega^3 & \omega^4\\
1 & \omega^4 & \omega^4 & 1 & \omega^2 & \omega^3 & \omega & \omega & \omega^3 & \omega^2\\\end{array}\right),
\end{equation}

where $\omega=e^{2\pi i/5}$ is a \emph{fifth} root of unity.

\begin{prop}
The matrix $S_{10}$ is inequivalent to $F^{(4)}_{10}$, $(F^{(4)}_{10})^T$, $D^{(3)}_{10}$, $D^{(8)}_{10}$, $N_{10A}$, $N^{(3)}_{10B}$, $G_{10}^{(1)}$ and $W'$.
\end{prop}
The matrix $S_{10}$ is found to be \emph{isolated} and contains only fifth roots of unity, therefore it is inequivalent to any of the complex Hadamard matrices listed in the proposition. However, we have not been able to show whether it is equivalent (or not) to the isolated  Butson-type matrix $X_{10} \in BH(10,5)$ or the matrix $B_{10}$ given in Appendix \ref{Butson's construction}.     

A different choice of the matrices $K_1$ and $L_1$, for example, $K_1=H_1$ and $L_1=H_2$, leads to the block matrix

\begin{equation}\label{d=10matrix}
S'_{10}=\frac{1}{\sqrt{2}}\left(\begin{array}{cc}
F_5 & H_2\\
H_1^{\dagger}F_5 & -H_1^{\dagger}H_2\end{array}\right),
\end{equation}
which in dephased form is given by 
\begin{equation}\label{dephasedmatrixd=10}
S'_{10}=\frac{1}{\sqrt{10}}\left(\begin{array}{cccccccccc}
1 & 1 & 1 & 1 & 1 & 1 & 1 & 1 & 1 & 1\\
1 & \omega & \omega^{2} & \omega^3 & \omega^4 & \omega^2 & \omega^3 & \omega^4 & 1 & \omega\\
1 & \omega^{2} & \omega^4 & \omega & \omega^3 & \omega^3 & 1 & \omega^2 & \omega^4 & \omega\\
1 & \omega^3 & \omega & \omega^4 & \omega^2 & \omega^3 & \omega & \omega^4 & \omega^2 & 1\\
1 & \omega^4 & \omega^3 & \omega^2 & \omega & \omega^2 & \omega & 1 & \omega^4 & \omega^3\\
1 & \omega^4 & \omega & \omega & \omega^4 & -1 & -\omega & -\omega^4 & -\omega^4 & -\omega\\
1 & \omega & 1 & \omega^2 & \omega^2 & -\omega^2 & -\omega & -\omega^2 & -1 & -1\\
1 & \omega^3 & \omega^4 & \omega^3 & 1 & -\omega^3 & -1 & -\omega^4 & -1 & -\omega^3\\
1 & 1 & \omega^3 & \omega^4 & \omega^3 & -\omega^3 & -\omega^3 & -1 & -\omega^4 & -1\\
1 & \omega^2 & \omega^2 & 1 & \omega & -\omega^2 & -1 & -1 & -\omega^2 & -\omega\\\end{array}\right),
\end{equation}
\normalsize
with $\omega=e^{2\pi i/5}$ a fifth root of unity. This matrix is a member of the family $BH(10,10)$ and, with a defect $d(S'_{10})=8$, the maximum dimension of any smooth manifold stemming from $S'_{10}$ will be eight. Several other matrices of the form $BH(10,10)$ exist but we are not able to determine whether $S'_{10}$ is  equivalent to any of them.
 
Any choice of $K_1$ and $L_1$ from the set of MU bases $\{F_5,H_1,H_2,H_3,H_4\}$  will result in a Butson-type matrix of the form $BH(10,5)$ or $BH(10,10)$. It would be interesting to see if the various combinations of $K_1$ and $L_1$ result in further new inequivalent complex Hadamard matrices.

\subsection{Dimension fourteen}

The known complex Hadamard matrices of order fourteen are the six-parameter Fourier family $F_{14}^{(6)}$ and its transpose $(F_{14}^{(6)})^T$, the family $D_{14}^{(5)}$ found in \cite{szollosi08}, and a set of isolated matrices $L_{14X}^{(0)}$ for $X=A,B,C,\ldots,N$ found in \cite{lampio12}. In addition, there are several Di\c{t}\v{a}-type matrices, listed in \cite{karol12}, obtained from Di\c{t}\v{a}'s method of Corollary \ref{thm:dita} using the Fourier matrix $F_2$ and any Hadamard matrix of order seven.

The matrix we construct from Theorem \ref{thm:newhadamard} consists of four blocks,
\begin{equation}
H_{14}=\frac{1}{\sqrt{2}}\left(\begin{array}{cc}
F_7 & L_1\\
K_1^{\dagger}F_7 & -K_1^{\dagger}L_1\end{array}\right),
\end{equation}
where the Fourier matrix of order seven is given by
\begin{equation}
F_{7}=\\\frac{1}{\sqrt{7}}\left(\begin{array}{ccccccc}
1 & 1 & 1 & 1 & 1 & 1 & 1\\
1 & \omega & \omega^{2} & \omega^3 & \omega^{4} & \omega^5 & \omega^{6}\\
1 & \omega^{2} & \omega^{4} & \omega^6 & \omega & \omega^3 & \omega^{5}\\
1 & \omega^3 & \omega^6 & \omega^{2} & \omega^5 & \omega & \omega^4 \\
1 & \omega^{4} & \omega & \omega^5 & \omega^2 & \omega^6 & \omega^3\\
1 & \omega^{5} & \omega^3 & \omega & \omega^{6} & \omega^4 & \omega^{2}\\
1 & \omega^6 & \omega^5 & \omega^4 & \omega^3 & \omega^{2} & \omega
\end{array}\right),
\end{equation}
with $\omega=e^{2\pi i/7}$, and the $(7\times 7)$ matrices $K_1$ and $L_1$ are chosen from the complete set of eight MU bases of the space $\mathbb{C}^7$.

Let us denote the complete set of MU bases by $\{I_7, F_7, H_1, H_2, H_3, H_4, H_5, H_6\}$ where
\begin{equation}
H_j=D^jF_7
\end{equation}
and
\begin{equation}
D=\text{diag}(1,1,\omega,\omega^3,\omega^6,\omega^3,\omega),
\end{equation}
with $j=1,\ldots,6$. The diagonal $D$ is based on the construction of a complete sets of MU bases in prime dimensions presented in \cite{bandyo02}. By choosing $K_1=H_1$ and $L_1=H_2$, we find a complex Hadamard matrix which, after dephasing, reads explicitly
\begin{equation}\label{isolatedmatrixd=15}%\nonumber
S_{14}=\small\frac{1}{\sqrt{14}}\left(\begin{array}{cccccccccccccc}
1 & 1 & 1 & 1 & 1 & 1 & 1 & 1 & 1 & 1 & 1 & 1 & 1 & 1 \\
1 & \omega & \omega^{2} & \omega^3 & \omega^{4} & \omega^5 & \omega^{6} & 1 & \omega & \omega^2 & \omega^{3} & \omega^4 & \omega^5 & \omega^6 \\
1 & \omega^{2} & \omega^{4} & \omega^6 & \omega & \omega^3 & \omega^{5} & \omega^2 & \omega^4 & \omega^6 & \omega & \omega^3 & \omega^5 & 1\\
1 & \omega^3 & \omega^6 & \omega^{2} & \omega^5 & \omega & \omega^4 & \omega^6 & \omega^2 & \omega^{5} & \omega & \omega^{4} & 1 & \omega^{3} \\
1 & \omega^{4} & \omega & \omega^5 & \omega^2 & \omega^6 & \omega^3 & \omega^{5} & \omega^{2} & \omega^{6} & \omega^{3} & 1 & \omega^{4} & \omega \\
1 & \omega^{5} & \omega^3 & \omega & \omega^{6} & \omega^4 & \omega^{2} & \omega^{6} & \omega^{4} & \omega^2 & 1 & \omega^{5} & \omega^{3} & \omega \\
1 & \omega^6 & \omega^5 & \omega^4 & \omega^3 & \omega^{2} & \omega & \omega^{2} & \omega & 1 & \omega^6 & \omega^{5} & \omega^4 & \omega^3 \\
1 & \omega & \omega^{3} & \omega^{6} & \omega^{3} & \omega & 1 & \omega^{5} & \omega^5 & \omega^{4} & \omega^2 & \omega^6 & \omega^{2} & \omega^{4}\\
1 & 1 & \omega & \omega^3 & \omega^6 & \omega^{3} & \omega & \omega^{4} & \omega^{5} & \omega^{5} & \omega^4 & \omega^2 & \omega^6 & \omega^2 \\
1 & \omega^6 & \omega^6 & 1 & \omega^2 & \omega^{5} & \omega^{2} & \omega & \omega^{3} & \omega^{4} & \omega^{4} & \omega^{3} & \omega & \omega^{5}\\
1 & \omega^5 & \omega^4 & \omega^4 & \omega^5 & 1 & \omega^{3} & \omega^{3} & \omega^6 & \omega & \omega^{2} & \omega^{2} & \omega & \omega^{6} \\
1 & \omega^{4} & \omega^2 & \omega & \omega & \omega^{2} & \omega^{4} & \omega^3 & 1 & \omega^{3} & \omega^{5} & \omega^{6} & \omega^{6} & \omega^{5}\\
1 & \omega^3 & 1 & \omega^5 & \omega^{4} & \omega^{4} & \omega^{5} & \omega & \omega^6 & \omega^{3} & \omega^{6} & \omega & \omega^{2} & \omega^{2}\\
1 & \omega^2 & \omega^5 & \omega^2 & 1 & \omega^{6} & \omega^6 & \omega^4 & \omega^{3} & \omega & \omega^{5} & \omega & \omega^{3} & \omega^{4} \normalsize
\end{array}\right),
\end{equation}
with $\omega=e^{2\pi i/7}$. This is an \emph{isolated} Butson-type complex Hadamard matrix of the form $BH(14,7)$.

\begin{prop}
The matrix $S_{14}$ is inequivalent to $F_{14}^{(6)}$, $(F_{14}^{(6)})^T$, $D_{14}^{(5)}$, $L_{14X}^{(0)}$ and every Di\c{t}\v{a}-type matrix given in \cite{karol12}. 
\end{prop}

Since $S_{14}$ and $L_{14X}^{(0)}$ contain different roots of unity they are inequivalent. All other known matrices listed in the proposition are contained in \emph{families} of complex Hadamard matrices, thus, they are inequivalent to $S_{14}$. However, it is not known whether $S_{14}$ is equivalent to the Butson-Hadamard matrix $B_{14}$ given in Appendix \ref{Butson's construction}.

We can construct further complex Hadamard matrices by choosing different MU bases for $K_1$ and $L_1$, e.g. if $K_1=H_1$ and $L_1=H_4$, the resulting matrix in dephased form is
\begin{equation}\label{isolatedmatrixd=15}%\nonumber
S'_{14}=\small\frac{1}{\sqrt{14}}\left(\begin{array}{cccccccccccccc}
1 & 1 & 1 & 1 & 1 & 1 & 1 & 1 & 1 & 1 & 1 & 1 & 1 & 1 \\
1 & \omega & \omega^{2} & \omega^3 & \omega^{4} & \omega^5 & \omega^{6} & 1 & \omega & \omega^2 & \omega^{3} & \omega^4 & \omega^5 & \omega^6 \\
1 & \omega^{2} & \omega^{4} & \omega^6 & \omega & \omega^3 & \omega^{5} & \omega^4 & \omega^6 & \omega & \omega^3 & \omega^5 & 1 & \omega^2\\
1 & \omega^3 & \omega^6 & \omega^{2} & \omega^5 & \omega & \omega^4 & \omega^5 & \omega & \omega^{4} & 1 & \omega^{3} & \omega^6 & \omega^{2} \\
1 & \omega^{4} & \omega & \omega^5 & \omega^2 & \omega^6 & \omega^3 & \omega^{3} & 1 & \omega^{4} & \omega & \omega^5 & \omega^{2} & \omega^6 \\
1 & \omega^{5} & \omega^3 & \omega & \omega^{6} & \omega^4 & \omega^{2} & \omega^{5} & \omega^{3} & \omega & \omega^6 & \omega^{4} & \omega^{2} & 1 \\
1 & \omega^6 & \omega^5 & \omega^4 & \omega^3 & \omega^{2} & \omega & \omega^{4} & \omega^3 & \omega^2 & \omega & 1 & \omega^6 & \omega^5 \\
1 & \omega & \omega^{3} & \omega^{6} & \omega^{3} & \omega & 1 & -\omega^{3} & -\omega & -\omega & -\omega^3 & -1 & -\omega^{6} & -1\\
1 & 1 & \omega & \omega^3 & \omega^6 & \omega^{3} & \omega & -1 & -\omega^{3} & -\omega & -\omega & -\omega^3 & -1 & -\omega^6 \\
1 & \omega^6 & \omega^6 & 1 & \omega^2 & \omega^{5} & \omega^{2} & -\omega^5 & -\omega^{6} & -\omega^{2} & -1 & -1 & -\omega^2 & -\omega^{6}\\
1 & \omega^5 & \omega^4 & \omega^4 & \omega^5 & 1 & \omega^{3} & -\omega^{4} & -\omega^3 & -\omega^4 & -1 & -\omega^{5} & -\omega^5 & -1 \\
1 & \omega^{4} & \omega^2 & \omega & \omega & \omega^{2} & \omega^{4} & -\omega^4 & -\omega & -1 & -\omega & -\omega^{4} & -\omega^{2} & -\omega^{2}\\
1 & \omega^3 & 1 & \omega^5 & \omega^{4} & \omega^{4} & \omega^{5} & -\omega^5 & -1 & -\omega^{4} & -\omega^{3} & -\omega^4 & -1 & -\omega^{5}\\
1 & \omega^2 & \omega^5 & \omega^2 & 1 & \omega^{6} & \omega^6 & -1 & -1 & -\omega^2 & -\omega^{6} & -\omega^5 & -\omega^{6} & -\omega^{2} \normalsize
\end{array}\right),
\end{equation}
with $\omega=e^{2\pi i/7}$. This is a Butson-Hadamard matrix of the form $BH(14,14)$, and has a defect of 12. It is unknown if $S'_{14}$ is equivalent to a $BH(14,14)$ matrix contained within an existing family of Hadamard matrices.

\subsection{Dimension fifteen}

The only known complex Hadamard matrices of order fifteen are the eight-parameter Fourier family $F^{(8)}_{15}$, stemming from the Fourier matrix $F_{15}\simeq F_3\otimes F_5\in BH(15,15)$, and the transposed Fourier family $(F^{(8)}_{15})^T$. To construct a new $(15\times15)$ complex Hadamard matrix by means of Theorem \ref{thm:newhadamard}, we use the block matrix
\begin{equation}
H_{15}=\frac{1}{\sqrt{3}}\left(\begin{array}{ccc}
F_5 & L_1 & L_2\\
K_1^{\dagger}F_5 & \alpha K_1^{\dagger}L_1 & \alpha^2K_1^{\dagger}L_2\\
K_2^{\dagger}F_5 & \alpha^2K_2^{\dagger}L_1 & \alpha K_2^{\dagger}L_2\end{array}\right),
\end{equation}
where $\alpha=e^{2\pi i/3}$ is a third root of unity, $I_5,F_5,K_1,K_2,L_1$ and $L_2$ are  $(5\times5)$ matrices including the identity matrix $I_5$ and the Fourier matrix $F_5$ defined in Eq. ($\ref{fourierd=5}$).
The set $\{I_5,K_1,K_2\}$ is MU to the set $\{F_5,L_1,L_2\}$.

If we use the \emph{complete} set of six MU bases of the space $\mathbb{C}^5$, $\{I_5,F_5,H_1,H_2,H_3,H_4\}$, corresponding to $K_1=H_1$, $K_2=H_2$, $L_1=H_3$ and $L_2=H_4$, as defined in Eq. ($\ref{MUcompletesetd=5}$), the resulting complex Hadamard matrix becomes
\begin{equation}
S_{15}=\frac{1}{\sqrt{3}}\left(\begin{array}{ccc}
F_5 & H_3 & H_4\\
H_1^{\dagger}F_5 & \alpha H_1^{\dagger}H_3 & \alpha^2H_1^{\dagger}H_4\\
H_2^{\dagger}F_5 & \alpha^2H_2^{\dagger}H_3 & \alpha H_2^{\dagger}H_4\end{array}\right).
\end{equation}
Apart from a factor $1/\sqrt{15}$, its dephased form reads explicitly
\begin{multline}\label{isolatedmatrixd=15}%\nonumber
\tiny\left(\begin{array}{ccccccccccccccc}
1 & 1 & 1 & 1 & 1 & 1 & 1 & 1 & 1 & 1 & 1 & 1 & 1 & 1 & 1\\
1 & \omega^3 & \omega^{6} & \omega^9 & \omega^{12} & \omega^9 & \omega^{12} & 1 & \omega^3 & \omega^6 & \omega^{12} & 1 & \omega^3 & \omega^6 & \omega^9\\
1 & \omega^{6} & \omega^{12} & \omega^3 & \omega^9 & \omega^6 & \omega^{12} & \omega^3 & \omega^9 & 1 & \omega^3 & \omega^9 & 1 & \omega^6 & \omega^{12}\\
1 & \omega^9 & \omega^3 & \omega^{12} & \omega^6 & \omega^6 & 1 & \omega^9 & \omega^3 & \omega^{12} & \omega^3 & \omega^{12} & \omega^6 & 1 & \omega^9\\
1 & \omega^{12} & \omega^9 & \omega^6 & \omega^3 & \omega^9 & \omega^6 & \omega^3 & 1 & \omega^{12} & \omega^{12} & \omega^9 & \omega^6 & \omega^3 & 1\\
1 & \omega^{12} & \omega^3 & \omega^3 & \omega^{12} & -\omega^5 & -\omega^{14} & -\omega^{11} & -\omega^{14} & -\omega & -\omega^{10} & -\omega & -\omega^4 & -\omega^4 & -\omega\\
1 & \omega^3 & 1 & \omega^6 & \omega^6 & -\omega^{2} & -\omega^8 & -\omega^{2} & -\omega^{14} & -\omega^{14} & -\omega^4 & -\omega^{13} & -\omega^4 & -\omega^7 & -\omega^7\\
1 & \omega^{9} & \omega^{12} & \omega^{9} & 1 & -\omega^{8} & -\omega^{11} & -\omega^2 & -\omega^{11} & -\omega^{8} & -\omega & -\omega^{13} & -\omega^{7} & -\omega^{13} & -\omega\\
1 & 1 & \omega^{9} & \omega^{12} & \omega^{9} & -\omega^{8} & -\omega^{8} & -\omega^{11} & -\omega^2 & -\omega^{11} & -\omega & -\omega & -\omega^{13} & -\omega^{7} & -\omega^{13}\\
1 & \omega^6 & \omega^6 & 1 & \omega^3 & -\omega^{2} & -\omega^{14} & -\omega^{14} & -\omega^{2} & -\omega^{8} & -\omega^4 & -\omega^7 & -\omega^7 & -\omega^4 & -\omega^{13}\\
1 & \omega^6 & \omega^9 & \omega^9 & \omega^6 & -\omega^{10} & -\omega^{13} & -\omega^{7} & -\omega^{7} & -\omega^{13} & \omega^{5} & \omega^{14} & \omega^{11} & \omega^{11} & \omega^{14}\\
1 & \omega^9 & 1 & \omega^3 & \omega^3 & -\omega^{7} & -\omega^{4} & -\omega^{7} & -\omega & -\omega & \omega^{8} & \omega^{14} & \omega^{8} & \omega^{5} & \omega^{5}\\
1 & \omega^{12} & \omega^6 & \omega^{12} & 1 & -\omega^{13} & -\omega^{4} & -\omega & -\omega^{4} & -\omega^{13} & \omega^{2} & \omega^{5} & \omega^{11} & \omega^{5} & \omega^{2}\\
1 & 1 & \omega^{12} & \omega^6 & \omega^12 & -\omega^{13} & -\omega^{13} & -\omega^{4} & -\omega & -\omega^{4} & \omega^{2} & \omega^{2} & \omega^{5} & \omega^{11} & \omega^{5}\\
1 & \omega^3 & \omega^3 & 1 & \omega^9 & -\omega^{7} & -\omega & -\omega & -\omega^{7} & -\omega^{4} & \omega^{8} & \omega^{5} & \omega^{5} & \omega^{8} & \omega^{14}
\normalsize
\end{array}\right),
\end{multline}
where $\omega=e^{2\pi i /15}$ is now a \emph{fifteenth} root of unity. Since all matrix elements can be written in terms of 30th roots of unity, $S_{15}$ is an example of a Butson-Hadamard matrix $BH(15,30)$. The vanishing defect of this matrix, i.e. $d(S_{15})=0$, implies that $S_{15}$ is \emph{isolated}. This property excludes $S_{15}$ from being a member of either of the affine families $F_{15}^{(8)}$ or $(F_{15}^{(8)})^T$.

\begin{prop}
The matrix $S_{15}$ is inequivalent to $F_{15}^{(8)}$ and $(F_{15}^{(8)})^T$.
\end{prop}

One could produce additional complex Hadamard matrices by choosing different combinations of MU bases from the complete set of six, such as $K_1=I_5$ or $L_1=F_3$. It is likely that various inequivalent matrices will result from these choices.

\section{Examples: dimensions $d>15$, and further generalisations}\label{sec:generalisations}

The construction of the matrices $S_6$, $S_9$ and $S_{15}$ has a common feature: in each case, the matrices $K_0,\ldots,K_{p-1},L_0,\ldots,L_{p-1}$ used to construct the blocks of the Hadamard matrix in Theorem \ref{thm:newhadamard} include a \emph{complete} set of $(q+1)$ MU bases of the space $\mathbb{C}^q$. For the cases $d=6$ and $d=15$, i.e. $q=2p-1$, complete sets of MU bases in dimension three and five are used, respectively, resulting in the isolated matrices $S_6$ and $S_{15}$. Furthermore, in the case $d=9$,  where $q<2p-1$, a complete set of four MU bases in dimension three is used, and again we find an isolated Hadamard matrix, namely $S_9$. 

Thus, one might expect additional isolated complex Hadamard matrices to emerge for larger composite dimensions whenever its factors are related by $q\leq2p-1$. We have been able to confirm this property for all primes $p,q$, with $pq<100$ and $q\leq2p-1$, excluding the case $d=4$. The first three examples are covered by $S_6$, $S_9$ and $S_{15}$ which we already know are isolated. The remaining five matrices $S_{25}$, $S_{35}$, $S_{49}$, $S_{77}$ and $S_{91}$, also turn out to be \emph{isolated}. We construct these matrices as follows:

$\bullet$ $S_{25}$ is derived from the complete set of six MU bases $I_5$, $F_5$ and $H_j=D^jF_5$, $j=1\ldots4$, where $D=\text{diag}(1,\omega,\omega^4,\omega^4,\omega)$ and $\omega=e^{2\pi i/5}$. The matrices $K_n$, for $n=0\ldots4$, are chosen as $I_5,I_5,I_5,H_1,H_2$, respectively, and $L_n$ as $F_5,F_5,F_5,H_3,H_4$, respectively.

$\bullet$ $S_{35}$ uses the complete set of eight MU bases $I_7$, $F_7$ and $H_j=D^jF_7$, $j=1\ldots6$, where $D=\text{diag}(1,1,\omega,\omega^3,\omega^6,\omega^3,\omega)$ and $\omega=e^{2\pi i/7}$. The matrices $K_n$, for $n=0\ldots4$, are chosen as $I_7,I_7,H_1,H_2,H_3$, and $L_n$ as $F_7,F_7,H_4,H_5,H_6$, respectively.

$\bullet$ $S_{49}$ is constructed from the same complete set of MU bases used for $S_{35}$, and we choose $K_n$ as $I_7,I_7,I_7,I_7,H_1,H_2,H_3$, and $L_n$ as $F_7,F_7,F_7,F_7,H_3,H_4,H_5$, with $n=0\ldots6$, respectively.

$\bullet$ $S_{77}$ uses a complete set of twelve MU bases given by $I_{11}$, $F_{11}$ and $H_j$, $j=1\ldots10$ with $D=\text{diag}(1,1,\omega,\omega^3,\omega^6,\omega^{10},\omega^4,\omega^{10},\omega^6,\omega^3,\omega)$ and $\omega=e^{2\pi i/11}$. The matrices $K_n$, for $n=0\ldots6$, are chosen as $I_{11},I_{11},H_1,H_2,H_3,H_4,H_5$ and $L_n$ as $F_{11},F_{11},H_6,H_7,H_8,H_9,H_{10}$, respectively. 

$\bullet$ $S_{91}$ is based on the complete set of fourteen MU bases in $\mathbb{C}^{13}$, i.e.  the identity $I_{13}$, the Fourier matrix $F_{13}$, and the matrices $H_j=D^jF_{13}$ for $j=1\ldots 12$, where the diagonal matrix is given by $D=\text{diag}(1,1,\omega,\omega^3,\omega^6,\omega^{10},\omega^2,\omega^8, \omega^2,\omega^{10},\omega^6,\omega^3,\omega)$ and $\omega=e^{2\pi i/13}$. The matrices $K_1,\ldots,K_6,L_1,\ldots,L_6$, correspond to $H_1,\ldots,H_{12}$, respectively.

All these isolated matrices are of Butson-Hadamard type with $S_{25}\in BH(25,10)$, $S_{35}\in BH(35,70)$, $S_{49}\in BH(49,14)$, $S_{77}\in BH(77,154)$ and $S_{91}\in BH(91,182)$. They may have smaller roots of unity if their matrix elements contain no entries equal to $(-1)$. The matrix $S_{91}$ is similar to $S_{6}$ and $S_{15}$ in the sense that the prime factors of $d=91$ satisfy the equality $q=2p-1$, meaning that each MU basis from the complete set is used exactly once. For the other isolated matrices, the factors satisfy the inequality $q<2p-1$, which implies that some MU bases are used more than once in the set $K_0,\ldots,L_{p-1}$. In this case, there are additional choices for the bases used; different combinations may lead to further inequivalent isolated complex Hadamard matrices.

So far, we have applied Theorem \ref{thm:newhadamard} mainly to product dimensions $4\leq d \leq 15$ or when $d=pq<100$ and $q\leq2p-1$. To explore whether the latter constraint on the factors $p$ and $q$ is necessary, we have constructed Hadamard matrices in all other composite dimension $d<100$ for $p,q\leq 13$. In each of these cases, i.e. $d=21$, 22, 26, 33, 39, 55 and 65, we were able to identify isolated Hadamard matrices. In addition, it is also possible to construct Hadamard matrices with non-zero defects, simply by selecting different sets of MU bases for $K_0,\ldots,L_{p-1}$. Thus, the theorem is potentially the source of infinitely many new Hadamard matrices in \emph{arbitrary} product dimensions.

Interestingly, the method is not limited to dimensions of the form $d=pq$: if the numbers $p$ and $q$ are composite, it is likely that additional, possibly inequivalent complex Hadamard matrices can be constructed which relate to different factorisations of the dimension, such as $2\times 6=3\times 4$ when $d=12$. Furthermore, it has been shown \cite{Kantor12} that \emph{inequivalent} complete sets of MU bases exist for large prime powers $d=p^n$, possibly leading to yet more inequivalent Hadamard matrices. One could also try to create continuous families of complex Hadamard matrices if sets of four or more MU bases exist which contain free parameters after dephasing.

Finally, another generalisation of Theorem \ref{thm:newhadamard} can be achieved as follows. The Hadamard matrices we have constructed are derived from product bases which tensor each vector in an orthonormal basis of $\mathbb{C}^p$ with an orthonormal basis of $\mathbb{C}^q$ (cf.  Eqs. (\ref{basis1},\ref{basis2})). However, other types of product bases exist; for example, one could take vectors from \emph{different} orthonormal bases in $\mathbb{C}^p$ and tensor them with vectors from \emph{one} basis in $\mathbb{C}^q$. The classification of all product bases in the space $\mathbb{C}^2\otimes\mathbb{C}^3$, up to local equivalence transformations, contains a number of examples of these so-called \emph{indirect} product bases \cite{mcnulty11}. Thus, alternative block structures may be allowed in Theorem \ref{thm:newhadamard}, potentially leading to other Hadamard matrices. 

\section{Summary and outlook \label{Sec:Summary}}

The main results of this paper are (i) a new general construction of complex Hadamard matrices in composite dimensions $d=pq$ ($p$, $q$ prime) described in Theorem \ref{thm:newhadamard}, and (ii) the explicit derivation of various new complex Hadamard matrices as a consequence of this theorem. The construction relies on the simple idea that a suitable unitary transformation maps a pair of MU \emph{product} bases to its standard form in which the vectors of one basis turn into the columns of a complex Hadamard matrix. It becomes possible to \emph{systematically} construct new Hadamard matrices many of which are isolated. Previous examples of isolated Hadamard matrices have been found by trial and error \cite{tao04} or from numerical methods \cite{beauchamp06}. 

To illustrate the approach we first derive some known results in low dimensions. In particular, we find the complete family of complex Hadamard matrices when $d=4$, and in dimension six we find the isolated matrix $S_6$. We then proceed to higher dimensions, obtaining isolated Hadamard matrices of order 9, 10, 14 and 15. Two of these are new isolated Butson-type Hadamard matrices, namely $S_{9}\in BH(9,6)$ and $S_{15}\in BH(15,30)$, while $S_{10}\in BH(10,5)$ and $S_{14}\in B(14,7)$ are shown to be inequivalent to nearly all known Hadamard matrices of their order. However, we cannot exclude the equivalence of $S_{10}$ to $B_{10}$ or $X_{10}$, and of $S_{14}$ to $B_{14}$.
 
In dimensions $d=10$ and $d=14$, there is some flexibility in selecting suitable subsets of MU bases when applying Theorem \ref{thm:newhadamard}. This enables us to construct two non-isolated Hadamard matrices $S'_{10}$ and $S'_{14}$, with defects equal to 8 and 12, respectively. Further research is needed to understand which choices of MU bases will lead to inequivalent Hadamard matrices.

Whenever the factors in the product dimension $d=pq$ are related by $q\leq2p-1$, the set $K_0,\ldots,K_{p-1},L_0$, $\ldots$, $L_{p-1}$ given in Theorem \ref{thm:newhadamard} can accommodate a \emph{complete} set of MU bases for the space $\mathbb{C}^q$. We speculate that in these cases, with the exception of dimension four, Theorem \ref{thm:newhadamard} will always give rise to an isolated Hadamard matrix. This expectation has been confirmed for all matrices of order $d=pq<100$ that satisfy $q\leq 2p-1$. In these cases, the matrices $S_6$, $S_9$, and $S_{15}$, as well as $S_{25}$, $S_{35}$, $S_{49}$, $S_{77}$ and $S_{91}$ all turn out to be isolated, and they include the largest known examples of isolated Hadamard matrices (as far as we know). What is more, we are also able to generate isolated and (non-isolated) Hadamard matrices for dimensions $d=21$, 22, 26, 33, 39, 55 and 65, giving rise to a total of 16 isolated complex Hadamard matrices. Twelve of them are new, while the remaining four, namely $S_{10}, S_{14}, S_{22}, S_{26}\in BH(2p,p)$, may be equivalent to matrices resulting from Butson's construction.  

Throughout this paper we have limited our search to Butson-type Hadamard matrices. However, the method given in Theorem \ref{thm:newhadamard} covers a much wider class of complex Hadamard matrices. We expect that many other examples of more general Hadamard matrices can be found by extending the choice for the unitary matrices $K_n$ and $L_n$.

\subsection*{Acknowledgments}

We would like to thank K. \.Zyczkowski and W. Tadej for sharing with us an efficient  program to calculate the defect of Hadamard matrices, and F. Sz\"{o}ll\H{o}si for helpful comments and suggestions. This work has been supported by EPSRC. 

\newpage
\appendix

\section{Appendix: Explicit construction of $BH(10,5)$ and $BH(14,7)$ \label{Butson's construction}}

In this Appendix we list the two Butson-type Hadamard matrices $BH(2p,p)$ of order 10 and 14, which we derive from the construction given in Butson's original paper \cite{butson62}.

$\bullet$ For $p=5$, the dephased matrix $BH(10,5)$ is given by

\begin{equation}
B_{10}=\frac{1}{\sqrt{10}}\left(\begin{array}{cccccccccc}
1 & 1 & 1 & 1 & 1 & 1 & 1 & 1 & 1 & 1\\
1 & \omega & \omega^{2} & \omega^3 & \omega^4 & \omega^2 & \omega^4 & \omega & \omega^3 & 1\\
1 & \omega^{2} & \omega^4 & \omega & \omega^3 & \omega^3 & \omega^2 & \omega & 1 & \omega^4\\
1 & \omega^3 & \omega & \omega^4 & \omega^2 & \omega^3 & \omega^4 & 1 & \omega & \omega^2\\
1 & \omega^4 & \omega^3 & \omega^2 & \omega & \omega^2 & 1 & \omega^3 & \omega & \omega^4\\
1 & \omega^3 & \omega^2 & \omega^2 & \omega^3 & 1 & \omega & \omega^4 & \omega^4 & \omega\\
1 & \omega^2 & 1 & \omega^4 & \omega^4 & \omega & \omega & \omega^3 & \omega^2 & \omega^3\\
1 & \omega & \omega^3 & \omega & 1 & \omega^4 & \omega^3 & \omega^4 & \omega^2 & \omega^2\\
1 & 1 & \omega & \omega^3 & \omega & \omega^4 & \omega^2 & \omega^2 & \omega^4 & \omega^3\\
1 & \omega^4 & \omega^4 & 1 & \omega^2 & \omega & \omega^3 & \omega^2 & \omega^3 & \omega\\\end{array}\right),
\end{equation}
where $\omega=e^{2\pi i/5}$ is a fifth root of unity. The defect of $B_{10}$ is zero.

$\bullet$ For $p=7$, the dephased matrix $BH(14,7)$ is

\begin{equation}\label{isolatedmatrixd=15}%\nonumber
B_{14}=\small\frac{1}{\sqrt{14}}\left(\begin{array}{cccccccccccccc}
1 & 1 & 1 & 1 & 1 & 1 & 1 & 1 & 1 & 1 & 1 & 1 & 1 & 1 \\
1 & \omega & \omega^{2} & \omega^3 & \omega^{4} & \omega^5 & \omega^{6} & \omega^6 & \omega^2 & \omega^5 & \omega & \omega^4 & 1 & \omega^3 \\
1 & \omega^{2} & \omega^{4} & \omega^6 & \omega & \omega^3 & \omega^{5} & \omega^3 & \omega^2 & \omega & 1 & \omega^6 & \omega^5 & \omega^4\\
1 & \omega^3 & \omega^6 & \omega^{2} & \omega^5 & \omega & \omega^4 & \omega^5 & 1 & \omega^{2} & \omega^4 & \omega^{6} & \omega & \omega^{3} \\
1 & \omega^{4} & \omega & \omega^5 & \omega^2 & \omega^6 & \omega^3 & \omega^{5} & \omega^{3} & \omega & \omega^{6} & \omega^4 & \omega^{2} & 1 \\
1 & \omega^{5} & \omega^3 & \omega & \omega^{6} & \omega^4 & \omega^{2} & \omega^{3} & \omega^{4} & \omega^5 & \omega^6 & 1 & \omega & \omega^2 \\
1 & \omega^6 & \omega^5 & \omega^4 & \omega^3 & \omega^{2} & \omega & \omega^{6} & \omega^3 & 1 & \omega^4 & \omega & \omega^5 & \omega^2 \\
1 & \omega^4 & \omega^{2} & \omega & \omega & \omega^2 & \omega^4 & 1 & \omega^5 & \omega^{6} & \omega^3 & \omega^3 & \omega^{6} & \omega^{5}\\
1 & \omega & \omega^3 & \omega^6 & \omega^3 & \omega & 1 & \omega^{4} & \omega^{6} & \omega^{4} & \omega^5 & \omega^2 & \omega^2 & \omega^5 \\
1 & \omega^5 & \omega^4 & \omega^4 & \omega^5 & 1 & \omega^{3} & \omega^2 & \omega & \omega^{3} & \omega & \omega^{2} & \omega^6 & \omega^{6}\\
1 & \omega^2 & \omega^5 & \omega^2 & 1 & \omega^6 & \omega^{6} & \omega & \omega^4 & \omega^3 & \omega^{5} & \omega^{3} & \omega^4 & \omega \\
1 & \omega^6 & \omega^6 & 1 & \omega^2 & \omega^5 & \omega^2 & \omega & \omega & \omega^{4} & \omega^{3} & \omega^{5} & \omega^{3} & \omega^{4}\\
1 & \omega^3 & 1 & \omega^5 & \omega^{4} & \omega^{4} & \omega^{5} & \omega^2 & \omega^6 & \omega^{6} & \omega^{2} & \omega & \omega^{3} & \omega\\
1 & 1 & \omega & \omega^3 & \omega^6 & \omega^{3} & \omega & \omega^4 & \omega^{5} & \omega^2 & \omega^{2} & \omega^5 & \omega^{4} & \omega^{6} \normalsize
\end{array}\right),
\end{equation}
with $\omega=e^{2\pi i/7}$ being a seventh root of unity. This matrix has zero defect.

\end{document}